\documentstyle[12pt]{article}

\newcommand{\be}{\begin{equation}}
\newcommand{\ee}{\end{equation}}
\newcommand{\Dlt}{\Delta}
\newcommand{\dlt}{\delta}

\newcommand{\bbe}{{\bf e}}
\newcommand{\bS}{{\bf S}}
\newcommand{\bB}{{\bf B}}

\newcommand{\al}{\alpha}
\newcommand{\ra}{\rightarrow}

\newcommand{\om}{\omega}
\newcommand{\Om}{\Omega}

\begin{document}

\begin{center}   

{\Large{\bf Adiabatic theorems for linear and nonlinear 
Hamiltonians} \\ [5mm]

V.I. Yukalov} \\ [3mm]

{\it Bogolubov Laboratory of Theoretical Physics, \\
 Joint Institute for Nuclear Research, Dubna 141980, Russia}

\end{center}

\vskip 3cm

\begin{abstract}

Conditions for the validity of the quantum adiabatic 
approximation are analyzed. For the case of linear Hamiltonians, 
a simple and general sufficient condition is derived, which is 
valid for arbitrary spectra and any kind of time variation. It 
is shown that in some cases the found condition is necessary and 
sufficient. The adiabatic theorem is generalized for the case of 
nonlinear Hamiltonians. 
\end{abstract}

\vskip 2cm


\newpage 

\section{Introduction}

The quantum adiabatic theorem is one of the principal results in 
quantum mechanics [1]. The standard consideration assumes a quantum  
system, whose Hamiltonian $H(t)$ varies slowly in time [1,2]. The 
wave function satisfies the Schr\"{o}dinger equation 
\be
\label{1}
i\dot{\psi}(t) = H(t) \; \psi(t) \; ,
\ee
where the overdot, as usual, implies time differentiation. Here 
and in what follows, the system of units is employed, where 
$\hbar=1$. For the sake of compactness, the matrix notation is 
employed, when the wave function is treated as a vector-column 
with respect to its spatial, spin, and other variables, except 
time $t$; and the Hamiltonian is a matrix in these variables. 
The wave function is normalized to one, $||\psi(t)|| = 1$, with 
the Euclidean vector norm assumed. The system Hamiltonian contains 
an explicit dependence on time.

Keeping time as a parameter, one considers the eigenproblem
\be
\label{2}   
H(t) \; \psi_n(t) = E_n(t) \; \psi_n(t) \; .
\ee
The eigenfunctions are normalized to one, $||\psi_n(t)|| = 1$.
The multi-index $n$, in general, can pertain to a discrete or 
continuous set.

If at the initial time $t = 0$ the system starts from a state
\be
\label{3}
\psi(0) = \psi_j(t) \; ,
\ee 
with a fixed $j$, and the Hamiltonian variation is sufficiently 
slow, then at the moment $t$ it is in a state that is close to 
$\psi_j(t)$. This, roughly speaking, is the meaning of the 
adiabatic theorem (see details in Refs. [1,2]). The criterion on 
the slowness, required by the theorem, is often formulated [3,4]
as the inequality
\be
\label{4}
\left |\; \frac{<\psi_n(t)\; | \; \dot{\psi}_j(t) > }{E_{nj}(t)}
\; \right | \; \ll \; 1 \qquad (n\neq j)
\ee
that is to be valid for all $n \neq j$, in the time interval 
$[0,\tau]$. Here
$$
E_{mn}(t) \equiv E_m(t) - E_n(t) \; .
$$ 
By differentiating Eq. (2), one gets 
$$
\dot{E}_n(t) \; = \; < \psi_n(t) \; | \; \dot{H}(t)\; | \;
\psi_n(t) > \; , 
$$
$$
E_{mn}(t) < \psi_m(t) \; | \; \dot{\psi}_n(t) > \; = \;
- < \psi_m(t) \; | \; \dot{H}(t)\; | \; \psi_n(t) > \; ,
$$
which allows to rewrite condition (4) in another form
\be
\label{5}
\left |\; \frac{<\psi_n(t)\;|\;\dot{H}(t)\;|\;\psi_j(t)>}{E_{nj}^2(t)} 
\; \right | \; \ll \; 1 \qquad (n\neq j) \; .
\ee
As is evident, conditions (4) or (5) have sense only when the 
denominator $E_{nj}\neq 0$. This imposes the restrictions on the 
spectrum of the considered system: The spectrum has to be 
nondegenerate; there should be no level crossings; and the 
fixed level $j$ has to be separated by a gap from other levels. 
Also, the requirement of the slow temporal variation of the 
Hamiltonian presupposes that the resonance case has to be 
excluded. This means that, if the effective frequency of the 
Hamiltonian variation is $\omega$, than it has to be smaller 
than any of the transition frequencies $E_{mn}(t)$ (see 
discussion in Ref. [6]). 

Moreover, conditions (4) or (5) are neither necessary nor 
sufficient. Thus, Messiah [4] mentions that such conditions are 
probably valid "in most cases". Since these conditions are not 
sufficient, their validity does not guarantee the applicability
of the adiabatic approximation. Therefore, when such a condition 
holds but the adiabatic approximation fails, there is no any 
inconsistency in the adiabatic theorem. This trivial fact can be 
illustrated by a number of examples [5-10] and checked 
experimentally [11].

There exist mathematically correct formulations of the adiabatic 
theorem, relaxing some of the requirements on the spectrum. Thus, 
Born and Fock [1] considered the case with level crossings. Avron 
and Elgart [12] proved the theorem without a gap condition. There 
have been considered the variants of the adiabatic theorem for 
open systems [13], in the presence of noise [14], as well as 
corrections to the adiabatic approximation [15]. This interest to 
formulating convenient conditions for the validity of the 
adiabatic approximation is supported by the recent discussions 
on the feasibility of adiabatic quantum computation [16,17].       

In the present paper, a novel simple and, at the same time, very  
general sufficient condition for the validity of the adiabatic 
approximation is derived. This condition is valid for arbitrary 
spectra and for any time variation of the system Hamiltonian, 
which is not required to be slow. If the latter is fast, the 
adiabatic approximation is respectively limited in time. By an 
explicit illustration, it is shown that, in some cases, the 
suggested condition is both necessary and sufficient. The second 
aim of the present paper is to generalize the adiabatic theorem 
to the case of nonlinear Hamiltonians. Such Hamiltonians are met, 
e.g., in nonlinear optics and in the physics of cold atoms (see 
review articles [18-26]).

\section{Linear Hamiltonians}

This is the standard case of quantum mechanics. The linear 
Hamiltonian $H(t)$ is self-adjoint. The set $\{\psi_n(t)\}$ of 
the eigenfunctions of Eq. (2) forms a complete orthonormal basis,
such that
$$
< \psi_m(t) \; | \; \psi_n(t) > \; = \; \dlt_{mn} \; .
$$
The solution to Eq. (1) can be expanded over this basis as 
\be
\label{6}
\psi(t) = \sum_n a_n(t) \exp\{ i\chi_n(t) \}\; \psi_n(t) \; ,
\ee
where the phase
\be
\label{7}
\chi_n(t) = \dlt_n(t) + \zeta_n(t) 
\ee
is the sum of the dynamic and geometric phases, respectively:
\be
\label{8}
\dlt_n(t) \equiv - \int_0^t E_n(t') \; dt' \; , \qquad
\zeta_n(t) \equiv i \int_0^t < \psi_n(t') \; | \;
\dot{\psi}_n(t') > dt' \; .
\ee
We may note that 
$$
< \dot{\psi}_m(t) \; | \; \psi_n(t)> \; = \;
- < \psi_m(t) \; | \; \dot{\psi}_n(t) > \; ,
$$
which follows from differentiating the ortonormality condition.
To be precise, let us give the definition of what will be called 
the adiabatic approximation.

{\bf Definition} ({\it Adiabatic approximation}). The function 
\be
\label{9}
\tilde\psi_j(t) =  a_j(t) \exp \{ i\chi_j(t) \}\; \psi_j(t)
\ee
is the adiabatic approximation for the solution $\psi(t)$ of 
Eq. (1), with the initial condition (3), if and only if
\be
\label{10}
|| \; \psi(t) - \tilde\psi_j(t)\; || \; \ll \; 1 \; ,
\ee
where the Euclidean vector norm is assumed. 

Inequality (10) tells us that the Euclidean distance between 
functions $\psi(t)$ and $\tilde\psi_j(t)$ is small. This 
inequality is the necessary and sufficient condition for  
function (9) to be called the adiabatic approximation. 

{\bf Theorem 1}. Let the Hamiltonian $H(t)$ be linear and 
self-adjoint, $\psi(t)$ be the solution to Eq. (1), with the 
initial condition (3), and let $\psi_n(t)$ be the solutions to 
eigenproblem (2), which are differentiable over time. Then 
function (9) is the adiabatic approximation, for the time 
interval $[0,\tau]$, in the sense of definition (10), if the 
condition
\be
\label{11}
\sum_{n(\neq j)} \; \int_0^t  
| < \psi_j(t') \; | \; \dot{\psi}_n(t') > | \; dt' \; \ll \; 1
\ee
holds for all $t \in [0,\tau]$.

{\bf Proof}. Substituting Eqs. (6) and (9) into the left-hand 
side of Eq. (10) gives
\be
\label{12}
||\; \psi(t) - \tilde\psi_j(t) \; ||^2 = 
 1  - |\; a_j(t)\; |^2 \; .
\ee
Hence, the necessary and sufficient condition (10) takes the 
form
\be
\label{13}
1 - |\; a_j(t) \; |^2 \; \ll \; 1 \; .
\ee
Using expansion (6) in Eq. (1) yields the equation
\be
\label{14}
\dot{a}_m(t) =  - \sum_{n(\neq m)} \; a_n(t) < \psi_m(t) \; | \;
\dot{\psi}_n(t) > \exp \{ - i\chi_{mn}(t) \} \; ,
\ee
where
\be
\label{15}
\chi_{mn}(t) \equiv \chi_m(t) - \chi_n(t) \; .
\ee
From Eq. (14), it follows that
\be
\label{16}
\frac{d}{dt} \; |\; a_m(t) \; |^2 =  - 2 {\rm Re}
\sum_{n(\neq m)} a_m^*(t) a_n(t) 
< \psi_m(t) \; | \; \dot{\psi}_n(t) > \exp\{ - i\chi_{mn}(t)\} \; .
\ee
Integrating the latter equation results in
$$
|\; a_m(t) \; |^2\;  =\;  | \; a_m(0) \; |^2 \; - 
$$
\be
\label{17}
- \; 2{\rm Re} 
\sum_{n(\neq m)} \; \int_0^t a_m^*(t') a_n(t')
< \psi_m(t') \; | \; \dot{\psi}_n(t') > 
\exp\{ - i\chi_{mn}(t')\} \; dt' \; .
\ee
In view of the initial condition (3), one has
\be
\label{18}
a_n(0) = \dlt_{nj}  \; .
\ee
Using this and setting $m = j$ in Eq. (17) gives
\be
\label{19}
1 - | \; a_j(t)\; |^2 = 2{\rm Re} 
\sum_{n(\neq j)} \; \int_0^t a_j^*(t') a_n(t')
< \psi_j(t') \; | \; \dot{\psi}_n(t') > 
\exp\{ - i\chi_{mj}(t')\} \; dt' \; .
\ee
The necessary and sufficient condition for the validity of the 
adiabatic approximation is that the right-hand side of Eq. (19) 
be small, in agreement with inequality (13). Majorizing this 
right-hand side, with taking into account that $|a_n(t)| \leq 1$, 
leads to
\be
\label{20}
1 - | \; a_j(t)\; |^2 \; \leq \; 2\sum_{n(\neq j)} \; \int_0^t
|\; < \psi_j(t') \; | \; \dot{\psi}_n(t') > \; | \; dt' \; .
\ee
Therefore, for inequality (13) to hold, it is sufficient that
condition (11) be valid. This concludes the proof. 

{\bf Remark}. The summation in above formulas is over the spectral 
multi-index $n$, whose nature can be arbitrary. If it pertains 
to a continuous set, then the summation should be understood as 
integration. That is, the theorem is valid for discrete as well 
as for continuous spectra. In proving the theorem, no restrictions 
have been imposed on the spectral properties. The spectrum can be 
arbitrary, whether discrete or continuous, nondegenerate or 
degenerate, gapful or gapless, without or with level crossings. 
The rate of the Hamiltonian temporal variation is also arbitrary, 
including the resonance case. So, the sufficient condition (11) 
seems to be more general than many other known sufficient conditions.  
The relation of this condition (11) to other conditions of close 
forms will be considered in detail in the concluding section 
Discussion.

\section{Nonlinear Hamiltonians}

Now, let us try to expand the adiabatic theorem to the case of 
nonlinear Hamiltonians, such as appear in the problem of cold 
atoms [18-26]. Let a nonlinear Hamiltonian
\be
\label{21}
H(t) =  H[\psi,\;t] = H[|\psi|,\; t ] \; ,
\ee
depending on $|\psi(t)|$, be gauge-invariant, such that it 
is invariant with respect to the gauge transformation
\be
\label{22}
H\left [ \psi e^{i\al},\; t\right ] = H [\psi,\; t] \; ,
\ee
where $\alpha$ is real. This type of Hamiltonians is typical of 
coherent systems corresponding to Bose-Einstein condensate. The 
function $\psi(t)$ is normalized to one and satisfies the same 
equation (1). A particular case of the latter could be, e.g., 
the Gross-Pitaevskii equation that is a nonlinear 
Schr\"{o}dinger equation. The same initial condition (3) is 
assumed. 

In the place of the eigenvalue problem (2), we have
\be
\label{23}
H[\psi_n,\; t] \; \psi_n(t) = E_n(t) \; \psi_n(t) \; .
\ee
The eigenfunctions can be chosen to be normalized to one. We 
look for the solution of Eq. (1) in the same form (6). Strictly 
speaking, a nonlinear equation can possess several types of 
solutions. But we can limit the consideration by the class of 
solutions representable in form (6), where
\be
\label{24}
| \; a_n(t)\; | \; \leq \; 1 \; .
\ee

An important difference of the nonlinear case, as compared to 
the linear one, is that the Hamiltonian $H[\psi_n,t]$ is not 
Hermitian since
$$
< \psi_m(t) \; | \; H[\psi_n,\; t]\psi_n(t) > \; \neq \;
< H[\psi_n,\; t]\psi_m(t) \; | \; \psi_n(t)> \; ,
$$
for $m \neq n$. As a consequence, the eigenfunctions $\psi_m(t)$ 
and $\psi_n(t)$ are not orthogonal for $m \neq n$. Also, the 
matrix elements of the quantity 
\be
\label{25}
\Dlt_n(t) \equiv H(t) - E_n(t)
\ee
are not zero, that is,
$$
< \psi_m(t) \; | \; \Dlt_n(t) \; | \; \psi_n(t) > \; 
\neq \; 0 \; .
$$
This essentially complicates the consideration and makes more 
cumbersome sufficient conditions for the validity of the 
adiabatic approximation.

{\bf Theorem 2}. Let a nonlinear gauge-invariant Hamiltonian be
defined by Eqs. (21) and (22) and let the eigenfunctions of the 
eigenproblem (23) be time-differentiable. Let function (6) be a 
solution to Eq. (1), under conditions (3) and (24). Then 
function (9) is the adiabatic approximation for the time 
interval $[0,\tau]$, in the sense of definition (10), provided 
that the following conditions hold:
\be
\label{26}
\sum_{n(\neq j)} \; | \;
< \psi_n(t) \; | \; \psi_j(t) > \; | \; \ll \; 1 \; ,
\ee
\be
\label{27}
\sum_{n(\neq j)} \; \int_0^t \; | \;
< \psi_n(t') \; | \; \dot{\psi}_j(t') > \; | \; dt' \; 
\ll \; 1 \; ,
\ee
\be
\label{28}
\sum_{n} \; \int_0^t \; | \;
< \psi_j(t') \; | \; \Dlt_j(t') \; | \; \psi_n(t') > \; | \; dt' \; 
\ll \; 1 \; ,
\ee
for all $t \in [0,\tau]$.

{\bf Proof}. The left-hand side of inequality (10) now reads as 
$$
||\; \psi(t) - \tilde\psi_j(t)\; || = 1  - |\; a_j(t) \;|^2 \; - 
$$
\be
\label{29}
- \; 2{\rm Re} 
\sum_{n(\neq j)} a_j^*(t) a_n(t) < \psi_j(t) \; | \; \psi_n(t) >
\exp\{ i\chi_{nj}(t) \} \; .
\ee
Majorizing the right-hand side of Eq. (29) gives
\be
\label{30}
|| \;\psi(t) - \tilde\psi_j(t) \; || \; \leq \; 
1  - | \; a_j(t) \; |^2 +
\sum_{n(\neq j)} |\; <  \psi_j(t) \; | \; \psi_n(t) > \; | \; .
\ee
From here, it is seen that for inequality (10) to be true, it is 
sufficient that condition (13) be valid together with 
condition (26). One may notice that 
$$
1  - | \; a_j(t)\; |^2 = \left ( 1 - |\; a_j(t)\;| \right )
\left ( 1 + |\; a_j(t)\;| \right ) \; \leq \; 
2 \left ( 1 - |\; a_j(t)\;| \right ) \; .
$$
Therefore, instead of condition (13), it is sufficient to have 
the inequality
\be
\label{31}
1 - |\; a_j(t)\; | \; \ll \; 1 \; .
\ee
From Eq. (1), with function (6), we obtain
$$
\dot{a}_m(t) =  - \sum_{n(\neq m)} \left [\; a_n(t)
< \psi_m(t) \; | \; \dot{\psi}_n(t) > + \; 
\dot{a}_n(t) < \psi_m(t) \; | \; \psi_n(t) > \; \right ]
\exp\{ - i\chi_{mn}(t) \} \; -
$$
\be
\label{32}
- \; i \sum_n a_n(t) < \psi_m(t) \; | \; \Dlt_n(t) \; | \; 
\psi_n(t)> \exp\{ - i\chi_{mn}(t) \} \; .
\ee
Due to the Hamiltonian gauge invariance (22), the eigenproblem 
(23) is invariant with respect to the gauge transformation 
$$
\psi_n(t) \; \ra \; \psi_n(t)\; e^{i\al_n(t)} \; ,
$$
where $\al_n(t)$ is real. This makes it possible to impose the 
Fock gauge calibration 
\be
\label{33}
< \psi_n(t) \; | \; \dot{\psi}_n(t) > \; = \; 0 \; .
\ee
Then phase (7) becomes
\be
\label{34}
\chi_n(t) = \dlt_n(t) = - \int_0^t E_n(t') \; dt' \; .
\ee
Denoting the right-hand side of Eq. (32) by $R(t)$, we write 
\be
\label{35}
\dot{a}_m(t) \equiv R(t) \; .
\ee
Integrating this equation, setting $m = j$, and using Eq. (18), 
we get
\be
\label{36}
a_j(t) - 1  =\int_0^t R(t') \; dt' \; .
\ee
Rewriting the latter equation as
$$
1 = a_j(t) - \int_0^t R(t')\; dt'
$$
and majorizing here the right-hand side, we find
\be
\label{37}
1 \; \leq \; |\; a_j(t)\;| + \int_0^t |\; R(t')\;|\; dt' \; .
\ee
This means that for the validity of Eq. (31), it is sufficient 
that the inequality
\be
\label{38}
\int_0^t |\; R(t') \;| \; dt' \; \ll \; 1 
\ee
be valid. Integrating the right-hand side of Eq. (32), we invoke 
the integration by parts in the term
$$
\int_0^t \dot{a}_n(t') < \psi_m(t') \; | \; \psi_n(t') >
\exp\{ -i\chi_{mn}(t') \} \; dt' \; = 
$$
$$
= \; a_n(t) < \psi_m(t) \; | \; \psi_n(t) > 
\exp\{ -i\chi_{mn}(t)\} \; - \;
a_n(0) < \psi_m(0) \; | \; \psi_n(0) > 
\exp\{ -i\chi_{mn}(0)\} \; -
$$
$$
- \int_0^t a_n(t') \left [ \; 
< \dot{\psi}_m(t') \; | \; \psi_n(t') > \;  + \;
< \psi_m(t') \; | \; \dot{\psi}_n(t')> +  \right.
$$
\be
\label{39}
+ \left. i E_{mn}(t') < \psi_m(t') \; | \; \psi_n(t')> \; \right ]
\exp\{ -i\chi_{mn}(t')\} \; dt' \; .
\ee
Then Eq. (37), with $R(t)$ given by the right-hand side of 
Eq. (32), yields
$$
1 - |\; a_j(t)\; | \; \leq \; \sum_{n(\neq j)} \left [\; | \;
< \psi_j(t) \; | \; \psi_n(t)> \; | \; + \;
|\; < \psi_j(0) \; | \; \psi_n(0)> \; | \; \right ] \; +
$$
\be
\label{40}
+ \; \sum_{n(\neq j)} \; \int_0^t \; | \; 
< \dot{\psi}_j(t') \; | \; \psi_n(t')>  \; | \; dt' \; + \;
 \sum_n \; \int_0^t \; | \; 
< \psi_j(t') \; | \; \Dlt_j(t') \; | \; \psi_n(t')>  \; |
\; dt' \; .
\ee
From here it becomes evident that inequality (31) holds, provided
that conditions (26), (27), and (28) are satisfied. This concludes 
the proof.

\section{Explicit illustration}

To illustrate how the derived sufficient conditions work, it is 
reasonable to consider a simple case for which all calculations 
could be done explicitly and exactly. To this end, let us study
the motion of a spin 1/2 in a rotating magnetic field [27]. The 
related magnetic moment is ${\bf m} = \mu_0 {\bf S}$, where 
${\bf S} = {\vec \sigma}/2$, and $\mu_0 = e g_L/2 m c$, with 
${\vec\sigma}$ being the Pauli vector matrix; $g_L$, Land\'{e}
factor; $c$, light velocity; and $e$, electric charge (for 
concreteness, taken to be positive). The Hamiltonian reads as 
\be
\label{41}
H(t) = - \mu_0 \bS\cdot\bB \; ,
\ee
with the rotating magnetic field
\be
\label{42}
\bB = (\; B\cos \om t\;)\; \bbe_x + 
(\; B\sin\om t \;)\; \bbe_y \; ,
\ee
which, without the loss of generality, can be assumed to rotate 
in the $x-y$ plane. The snapshot eigenvalues of Hamiltonian (41) 
are
\be
\label{43}
E_1 = - \; \frac{1}{2}\; \om_0 \; , \qquad 
E_2 = \frac{1}{2} \; \om_0 \; ,
\ee
where 
\be
\label{44}
\om_0 \equiv \mu_0 B
\ee  
is the Larmor frequency. The snapshot eigenfunctions are
\begin{eqnarray}
\nonumber
\psi_1(t) = \frac{1}{\sqrt{2}} \left ( 
\begin{array}{c} 1 \\
                 0 \end{array} \right )  + 
\frac{e^{i\om t}}{\sqrt{2}} \left ( \begin{array}{c}
0 \\
1 \end{array} \right ) \; ,
\end{eqnarray}
\begin{eqnarray}
\label{45}
\psi_2(t) = \frac{1}{\sqrt{2}} \left ( 
\begin{array}{c} 1 \\
                 0 \end{array} \right )  -
\frac{e^{i\om t}}{\sqrt{2}} \left ( \begin{array}{c}
0 \\
1 \end{array} \right ) \; .
\end{eqnarray}

Condition (11) tells us that, if the system starts from the 
state $\psi_1(0)$, then its wave function $\psi(t)$ is close to 
the adiabatic approximation (9), for the time interval 
$[0,\tau]$, in the sense of Eq. (10), provided that
\be
\label{46}
\int_0^t | \; < \psi_1(t') \; | \; \dot{\psi}_2(t') > \; | \;
 dt' \; \ll \; 1 \; .
\ee
Similarly, when the system starts from $\psi_2(0)$, then the 
sufficient condition (11) becomes
\be
\label{47}
\int_0^t | \; < \psi_2(t') \; | \; \dot{\psi}_1(t') > \; | \;
 dt' \; \ll \; 1 \; .
\ee
From Eqs. (45), we have
\begin{eqnarray}
\label{48}
\dot{\psi}_1(t) = \frac{i}{\sqrt{2}} \; \om e^{i\om t} \left ( 
\begin{array}{c} 0 \\
                 1 \end{array} \right )  \; , \qquad
\dot{\psi}_2(t) = -\; \frac{i}{\sqrt{2}}\; \om e^{i\om t} \left ( 
\begin{array}{c} 0 \\
                 1 \end{array} \right )  \; .
\end{eqnarray}
Hence
\be
\label{49}
< \psi_1(t) \; | \; \dot{\psi}_2(t) > \; = \; 
< \psi_2(t) \; | \; \dot{\psi}_1(t)> \; = 
\; - \frac{i}{2} \; \om \; .
\ee
Thus, both Eqs. (46) and (47) result in the condition
\be
\label{50}
\om t \; \ll \; 1 \qquad (t\in[0,\tau]) \; .
\ee

This condition has two different representations, depending on 
the relation between the frequencies $\omega$ and $\omega_0$. 
If $1 / \omega_0 \leq \tau$, then there exists such a time 
$t = 1 / \omega_0$ which lies inside the interval $[0,\tau]$, 
which means that $\omega \ll \omega_0$. This case is 
summarized in the form of the conditions
\be
\label{51}
\om \; \ll \; \om_0 \; , \qquad \om\tau \; \ll \; 1 \; , \qquad
\om_0\tau \; \geq \; 1 \; .
\ee
The opposite situation is when $\omega$ is larger or of the 
order of $\omega_0$. Then Eq. (50) is equivalent to the 
inequalities
\be
\label{52}
\om \; \geq \; \om_0 \; , \qquad \om\tau \; \ll \; 1 \; , \qquad
\om_0\tau \; \ll \; 1 \; .
\ee

The first case (51) is in line with the standard understanding 
of the slow Hamiltonian variation, allowing for the use of the 
adiabatic approximation. But the second situation (52) corresponds 
to fast variation, including the resonance case, when $\omega$ 
coincides with $\omega_0$. For both these variants, there exists 
a limitation on the admissible time interval, during which 
the adiabatic approximation is applicable. The basic difference 
is that this time interval can be larger for the slow Hamiltonian  
variation, as compared to its fast resonance variation.   

Since the considered problem allows for an exact solution, we can 
check whether the found conditions are really sufficient. 
Equation (1), under the initial condition
\begin{eqnarray}
\label{53}
\psi(0) = c_1 \left ( 
\begin{array}{c} 1 \\
                 0 \end{array} \right )  +
c_2 \left ( \begin{array}{c}
0 \\
1 \end{array} \right ) \; ,
\end{eqnarray}
for Hamiltonian (41), results in the solution
\begin{eqnarray}
\label{54}
\psi(t) = b_1(t) \left ( 
\begin{array}{c} 1 \\
                 0 \end{array} \right )  +
b_2(t) \left ( \begin{array}{c}
0 \\
1 \end{array} \right ) \; ,
\end{eqnarray}
in which the coefficient functions are
$$
b_1(t) = \left (\; c_1 \cos\; \frac{\Om t}{2} + i \; 
\frac{c_1\om+c_2\om_0}{\Om}\;
\sin\; \frac{\Om t}{2} \; \right ) \; e^{-i\om t/2} \; ,
$$
\be
\label{55}
b_2(t) = \left (\; c_2 \cos\; \frac{\Om t}{2}  - i \; 
\frac{c_2\om-c_1\om_0}{\Om}\;
\sin\; \frac{\Om t}{2} \; \right )\; e^{i\om t/2} \; ,
\ee
and the notation is used for the effective Rabi frequency
\be
\label{56}
\Om \; \equiv \; \sqrt{\om^2 +\om_0^2} \; .
\ee
Inverting relations (45) gives
\begin{eqnarray}
\nonumber
\left (\begin{array}{c} 
1 \\
0 \end{array} \right ) = \frac{1}{\sqrt{2}} \; \psi_1(t) + 
\frac{1}{\sqrt{2}}\; \psi_2(t) \; ,
\end{eqnarray}
\begin{eqnarray}
\label{57}
\left (\begin{array}{c} 
0 \\
1 \end{array} \right ) = \frac{e^{-i\om t}}{\sqrt{2}} \; \psi_1(t) - 
\frac{e^{-i\om t}}{\sqrt{2}}\; \psi_2(t) \; .
\end{eqnarray}
So that expansion (6) over eigenfunctions (45) takes the form
\be
\label{58}
\psi(t) =  a_1(t)\; \psi_1(t) + a_2(t) \; \psi_2(t) \; ,
\ee
where
\be
\label{59}
a_1(t) = \frac{b_1(t)+b_2(t)}{\sqrt{2}} \; , \qquad 
a_2(t) = \frac{b_1(t)-b_2(t)}{\sqrt{2}} \; .
\ee
At the initial moment of time, we have
$$
b_1(0) = c_1 \; , \qquad b_2(0) = c_2 \; , \qquad
a_1(0) = \frac{c_1+c_2}{\sqrt{2}} \; , \qquad
a_2(0) = \frac{c_1-c_2}{\sqrt{2}} \; .
$$

Suppose, first, that the system starts from the eigenstate 
$\psi_1(0)$, when
\begin{eqnarray}
\label{60}
\psi(0) = \psi_1(0) = \frac{1}{\sqrt{2}} \left ( 
\begin{array}{c} 1 \\
                 0 \end{array} \right )  +
\frac{1}{\sqrt{2}} \left ( \begin{array}{c}
0 \\
1 \end{array} \right ) \; ,
\end{eqnarray}
so that
$$
a_1(0) = 1 \; , \qquad a_2(0) = 0 \; , \qquad 
c_1=c_2 = \frac{1}{\sqrt{2}} \; .
$$
Then Eqs. (55) yield
$$
b_1(t) = \left (\; \cos\; \frac{\Om t}{2} + 
i \; \frac{\om+\om_0}{\Om} \; \sin\; \frac{\Om t}{2}\; 
\right ) \; \frac{e^{-i\om t/2}}{\sqrt{2}} \; ,
$$
\be
\label{61}
b_2(t) = \left ( \; \cos\; \frac{\Om t}{2} \; -  \;
i \; \frac{\om-\om_0}{\Om} \; \sin\; \frac{\Om t}{2}\; 
\right ) \; \frac{e^{i\om t/2}}{\sqrt{2}} \; .
\ee
In order that the adiabatic approximation be valid for a time 
interval $[0,\tau]$, according to definition (10), it is 
necessary and sufficient that
\be
\label{62}
|\; a_2(t)\; |\; \ll \; 1 \; , \qquad 
1  - |\; a_1(t)\; | \; \ll \;  1
\ee
for this interval of time. From Eq. (59), we find
$$
|\; a_2(t)\; |^2 = \frac{1}{4} \left |\; ( 1-\cos\om t) 
\cos\; \frac{\Om t}{2} \; - \; \frac{\om-\om_0}{\Om} \; 
\sin\; \frac{\Om t}{2} \; \sin \om t \; \right |^2 \; + 
$$
\be
\label{63}
+ \; \frac{1}{4} \left | \; \left ( \frac{\om+\om_0}{\Om} + 
\frac{\om-\om_0}{\Om}\; \cos \om t \right ) \;
 \sin\; \frac{\Om t}{2} \; - \; 
\cos \; \frac{\Om t}{2} \; \sin \om t \; \right |^2 \; .
\ee
When inequality (50) holds, then Eq. (63) simplifies to 
\be
\label{64}
|\; a_2(t)\; | \simeq \left | \; \frac{\om}{\Om} \; 
\sin \; \frac{\Om t}{2} \; \right | \; .
\ee
As is explained above, Eq. (50) is equivalent to one of 
the conditions, either (51) or (52). In both the cases, 
expression (64) satisfies inequality (62). Conversely, from 
Eqs. (62) and (63), one gets inequality (50). That is, 
inequality (50) is the necessary and sufficient condition for 
the validity of the adiabatic approximation.

Let now the system start from the eigenstate $\psi_2(0)$, which 
means that 
\begin{eqnarray}
\label{65}
\psi(0) = \psi_2(0) =\frac{1}{\sqrt{2}} \left (
\begin{array}{c}
1 \\
0 \end{array} \right ) - \frac{1}{\sqrt{2}} \left (
\begin{array}{c}
0 \\
1 \end{array} \right ) \; ,
\end{eqnarray}
with
$$
a_1(0) = 0 \; , \qquad a_2(0) = 1 \; , \qquad 
c_1 = - c_2 = \frac{1}{\sqrt{2}} \; .
$$
Then one has
$$
b_1(t) = \left ( \; \cos\; \frac{\Om t}{2} + i \; 
\frac{\om-\om_0}{\Om} \; \sin \; \frac{\Om t}{2} \; \right ) \; 
\frac{e^{-i\om t/2}}{\sqrt{2}} \; ,
$$
\be
\label{66}
b_2(t) = \left ( \; - \cos\; \frac{\Om t}{2} + i \; 
\frac{\om+\om_0}{\Om} \; \sin \; \frac{\Om t}{2}\; \right ) \; 
\frac{e^{i\om t/2}}{\sqrt{2}} \; .
\ee
For the adiabatic approximation to hold, in the sense of 
definition (10), it is necessary and sufficient that 
\be
\label{67}
| \;a_1(t)\; |\; \ll \; 1 \; , \qquad 
1 - |\; a_2(t)\; | \; \ll \; 1 \; .
\ee
In view of Eqs. (59), we have
$$
|\; a_1(t)\; |^2 = \frac{1}{4} \left |\; ( 1-\cos\om t) 
\cos\; \frac{\Om t}{2} \; - \; \frac{\om+\om_0}{\Om} \; 
\sin\; \frac{\Om t}{2} \; \sin \om t \; \right |^2 +
$$
\be
\label{68}
+ \frac{1}{4} \left | \; \left ( \frac{\om- \om_0}{\Om} + 
\frac{\om+\om_0}{\Om} \cos \om t \right ) \;
\sin \; \frac{\Om t}{2}  - 
\cos\; \frac{\Om t}{2} \; \sin \om t \; \right |^2 \; .
\ee
Following the same reasoning as above, we see that under
inequality (50), the value of Eq. (68) becomes
\be
\label{69}
| \; a_1(t)\; | \simeq \left | \; \frac{\om}{\Om} \; \sin \;
\frac{\Om t}{2} \; \right | \; ,
\ee
which satisfies Eq. (67), because of either condition (51) 
or condition (52). Hence, again, inequality (50) is the necessary 
and sufficient condition for the applicability of the adiabatic 
approximation.

These examples demonstrate that inequality (11), found as a 
sufficient condition for the validity of the adiabatic 
approximation, in some cases, becomes the necessary and sufficient 
condition.     

At the same time, the standard estimate (4) for the considered 
case gives
\be
\label{70}
\left | \; \frac{ < \psi_2(t)\; | \;\dot{\psi}_1(t)>}{E_{21}(t)} \;
\right | = \frac{\om}{2\om_0}\; \ll \; 1 \; .
\ee
As is evident from the above consideration, this is neither 
necessary nor sufficient condition for the occurrence of the 
adiabatic approximation. This is why the situation, when 
inequality (70) is valid, but the adiabatic approximation is not 
correct, in no sense implies an inconsistency of the adiabatic 
theorem.

\section{Discussion}

A simple and general sufficient condition is derived for the 
validity of the adiabatic approximation in the case of quantum 
linear systems with arbitrary spectra and with any time variation 
of the related linear Hamiltonian. The adiabatic theorem is extended 
to the case of quantum nonlinear Hamiltonians. By considering an 
exactly solvable model, it is shown that the found sufficient 
conditions in some cases become necessary and sufficient. Contrary
to this, the usually considered condition (4) is shown to be neither 
necessary nor sufficient.  

It is important that the found conditions can be used for any type 
of the system spectrum, whether discrete or continuous, nondegenerate 
or degenerate, gapful or gapless, without or with level crossings. 
The temporal Hamiltonian variation can also be arbitrary, whether 
slow or fast, including the case of resonance transitions. The main 
difference between the slow and fast Hamiltonian variations is 
that in the former case the time interval, allowing for the validity 
of the adiabatic approximation, is larger than in the case of fast 
variation, when such a time interval can become rather short.

The summation over the spectral label $n$, appearing in formulas, 
can be explicitly realized for each particular system. There exist 
a variety of methods, both analytical and numerical, for evaluating 
such series. Moreover, for a great number of physical systems, 
only several lowest energy levels are of importance, which makes 
such systems effectively  spectrally bounded. Therefore, the 
arising summation does not lead to principal difficulties, but is 
rather a technical problem, which can be appropriately addressed 
for each particular case.   

To emphasize the difference of condition (11) from other 
sufficient conditions of close forms, it is worth mentioning 
some previous results. Thus, Tong et al. [10] considered a slow 
nonresonant Hamiltonian variation, for a finite-level system, 
with nondegenerate spectrum, without level crossings, and with a gap, 
separating the initial energy level from other levels. Under these 
restrictions, the formulated conditions are given by the set of 
the inequalities
$$
\left | \frac{< \psi_j(t)| \dot{\psi}_n(t)>}{E_{jn}(t)}
\right | \ll 1 \; ,
$$
$$
\int_0^\tau \left | \frac{d}{dt} \left [ 
\frac{< \psi_j(t)| \dot{\psi}_n(t)>}{E_{jn}(t)} \right ]
\right | \; dt \ll 1 \; ,
$$
$$
\int_0^\tau \left | 
\frac{<\psi_j(t) | \dot{\psi}_n(t)><\psi_n(t) | \dot{\psi}_m(t)>} 
{E_{jn}(t)}
\right | \; dt \ll 1 \; ,
$$
to be satisfied for any $t \in [0,\tau]$, all $n \neq j$, and any 
$m \neq n$. As is evident, these conditions ar quite different, 
and are accompanied by several restrictions that are not required 
for Theorem 1.  

Wei and Ying [28] showed that the Tong et al. conditions are valid, 
under the same restrictions plus the additional requirement that 
the Hamiltonian be real, if the following two inequalities hold:
$$
\left |
\frac{< \psi_j(t)| \dot{\psi}_n(t)>}{E_{jn}(t)} 
\right |  \ll 1 \; ,
$$
$$
\sum_{n(\neq j)} \int_0^\tau | < \psi_j(t) | \dot{\psi}_n(t)> | \;
dt \;  <  \; const \; ,
$$
for all $n \neq j$ and $t \in [0,\tau]$. Again, the main difference 
is that condition (11) does not require all these restrictions, thus, 
being essentially more general.

Maamache and Saadi [29,30] studied a slowly varying Hamiltonian, 
producing no resonant transitions and enjoying a continuous 
nondegenerate spectrum $E_k(t)$, having no level crossings, with 
the parameter $k$ inside a wave-packet range $[k_0, k_0 + \delta k_0]$,
where $|\delta k_0| \ll |k|$. Their result reads as
$$
\left | \int_0^\tau \exp \left \{ i \int_0^t \left [
E_{k_0}(t') - E_k(t') \right ] \; dt' \right \} 
< \psi_{k_0}(t) | \dot{\psi}_k(t)> dt \right |^2 \ll 1 \; ,
$$
$$
k \in [k_0, k_0+\dlt k_0 ] \; , \qquad | \dlt k_0 | \ll | k | \; .
$$
A similar condition can be obtained from Eqs. (13) and (19) for 
the case of the continuous spectrum. However, again, we do not 
need the restrictions that the Hamiltonian be slowly varying and 
nonresonant, and the spectrum is not required to be nondegenerate 
and without level crossings.  

For nonlinear systems, the applicability of the adiabatic 
approximation is substantially limited. The nonlinearity acts 
as an additional perturbation destroying the adiabatic evolution. 
If the nonlinearity is large, then the adiabatic approximation 
is limited by a short time interval $\tau$, as follows from the 
sufficient conditions (26) to (28). When the nonlinearity is 
small, such that the eigenfunctions of eigenproblem (23) are 
almost orthogonal between each other, then the problem reduces 
to the case of linear Hamiltonians.

Generally speaking, nonlinear quantum systems are not good 
candidates for the use of the adiabatic approximation that is 
strongly destroyed by nonlinearity. Therefore, when considering 
possible physical setups for adiabatic quantum computation 
[16,17], it is better to keep in mind noninteracting or at least 
very weakly interacting systems. As examples, one could take 
weakly interacting trapped atoms [18-26] or cold molecules [31]. 
Another physical system, convenient for quantum information 
processing, is a collection of spins. But again, the spins 
should not strongly interact with each other. This imposes a 
severe restriction on the density of spins, since they interact 
through dipole forces that are long-ranged and strongly influence 
the spin motion [32-34]. The treatment of nonlinear systems is 
essentially more complicated than that  of linear systems and 
requires a separate investigation for each particular case. For 
example, the validity of the adiabatic approximation for cold 
trapped atoms, subject to the action of an alternating trap 
modulation, is analyzed in Ref. [35]. The sufficient conditions, 
derived in the present paper, can be used for estimating the 
parameters of physical systems that are intended to be employed 
for realizing adiabatic processes, such as adiabatic quantum 
computing.
           
\vskip 5mm

{\bf Acknowledgement} 

\vskip 2mm

Financial support from the Russian Foundation for Basic Research 
(Grant 08-02-00118) is appreciated.

\end{document}